\documentclass{nature}

\usepackage{amsmath,amssymb,amsfonts,latexsym}
\usepackage[pdftex]{graphicx}

\title{Spontaneous synchrony in power-grid networks}

\author{
Adilson E. Motter$^{1,2,*}$,
Seth A. Myers$^{3}$,
Marian Anghel$^{4}$ and
Takashi Nishikawa$^{1}$
}

\newcounter{firstbib}

\begin{document}

\baselineskip18pt
\setlength{\parskip}{0.15in}

\maketitle

\noindent
\framebox{\emph{Nature Physics} {\bf 9}, 191--197 (2013) $|$ doi:10.1038/nphys2535}

\begin{affiliations}
 \item Department of Physics and Astronomy, Northwestern University, Evanston, Illinois 60208, USA.
 \item Northwestern Institute on Complex Systems, Northwestern University, Evanston, Illinois 60208, USA.
 \item Institute for Computational and Mathematical Engineering, Stanford University, Stanford, California 94305, USA.
 \item Information Sciences Group, Los Alamos National Laboratory, Los Alamos, New Mexico 87544, USA.
\end{affiliations}

\vspace{-7mm}\noindent $^*$ E-mail: \texttt{motter@northwestern.edu}.
\vspace{5mm}

\begin{abstract}
An imperative condition for the functioning of a power-grid network is that its power generators remain synchronized. 
Disturbances 
can prompt desynchronization, 
which is a process that has been involved in large power outages.
Here we derive a condition under which the desired synchronous state of a power grid is stable, and use this condition to identify tunable parameters of the generators that are determinants of spontaneous synchronization.  Our analysis gives rise to an approach to specify parameter assignments that can enhance synchronization of any given network, which we demonstrate for a selection of both test systems and real power grids. 
Because our results concern spontaneous synchronization, they are relevant both for reducing dependence on conventional control devices, thus offering an additional layer of protection given that most power outages involve equipment or operational errors, and for contributing to the development of ``smart grids'' that can recover from failures in real time.
\end{abstract}

\pagebreak

The current resounding interest in network synchronization\cite{review1,review2} has been stimulated by the prospect that theoretical
studies will help explain the behavior of real complex networks\cite{strogatz}. 
Recent advances include the  modeling  of chimera spatiotemporal patterns\cite{abrams2004},  the discovery of low-dimensional dynamics in heterogeneous
populations\cite{ott2008},  and the characterization of network synchronization landscapes\cite{nishikawa2010}.
However, despite the significant insights provided by these and other studies\cite{Hagerstrom2012,Tinsley2012,assenza2011,ravoori2011,hunt2010,sun2009,yu2009,kiss2007,retrepo2006}, 
the connection between network theory and the synchronization of real systems remains 
under-explored.
This is partly due to the scarcity of dynamical data, partly due to the idealized nature of the theoretical
constructs (for network-unrelated exceptions, see refs.\ \citen{global1} and \citen{global2}). 
In light of these considerations, we explore power-grid systems as genuine complex networks of broad
significance that are amenable to theoretical modeling and whose dynamics can be simulated reliably. As the largest 
man-made machines in existence, modern power-grid networks often consist of thousands of power substations and generators 
linked across thousands of kilometers. This complexity is reflected in a rich variety of collective behaviors and 
instabilities these systems reportedly exhibit\cite{gellings}. 

Here, we focus on the synchronization of power generators---a fascinating phenomenon that can occur  
spontaneously, that keeps all connected generators in pace, and whose failure constitutes 
an important source
 of instabilities in power-grid systems. 
In a network of $n$ alternating current generators, a synchronous state is characterized by
\begin{equation}\label{eq0}
\dot\delta_1 = \dot\delta_2=\dots= \dot\delta_n,
\end{equation}
where $\delta_i = \delta_i(t)$ represents the rotational phase of the $i$th generator and the dot represents the time derivative.  This synchronization frequency is assumed to be close (albeit not necessarily equal) to a reference frequency.
That coupled power generators can  synchronize spontaneously is well known, as popularized in 
ref.\ \citen{strogatz2003}, and this 
has generated recent interest in the physics community\cite{Lozano:2012qy,PhysRevLett.109.064101}.
 However, the governing factors and hence the extent to which this phenomenon may occur in real power grids remain 
essentially unaddressed. 
Among the other studies that have provided substantive insights into power grid synchronization we mention 
the analysis of nonlinear modes associated with 
instabilities (refs.\ \citen{susuki2011,Susuki:2009vn,Parrilo:1999fk} and references therein) and of an equivalence between power systems and networks of nonuniform Kuramoto 
oscillators\cite{dorfler2010,Dorfler:2011fk,fd-fb:09z}.
The former studies provide an informative decomposition of nonlinear oscillations that arises when the network loses synchrony.
The latter studies establish a sufficient condition for a power grid to converge to a synchronous state, which is obtained via singular perturbation analysis
 from a similar condition for nonuniform Kuramoto oscillators\cite{dorfler2010,fd-fb:09z}.
This condition 
is necessary and sufficient
for non-identical dynamical units under the assumption that the coupling strengths are uniform\cite{Dorfler:2011fk}.
In contrast, the stability condition we derive below is necessary and sufficient for heterogeneously coupled power grids, with the assumption that a certain 
function of  generator parameters
 is homogeneous.
This condition helps us address the question of how to strengthen synchronization.

Power grids deliver a growing share of the energy consumed in the world and will soon undergo substantial changes owing to the increased harnessing of intermittent energy sources, the commercialization of plug-in electric automobiles, and the development of real-time pricing and two-way energy exchange technologies. These advances will further increase the economical and societal importance of power grids, but they will also lead to new disturbances associated with fluctuations in production and demand, which may trigger desynchronization of power generators.
Although power grids can, and often do, rely on active control devices to maintain synchronism, the question of whether proper design would allow the same to be achieved 
while reducing dependence on existing
controllers is extremely relevant. In the U.S., for example, data on reported power outages show that over 3/4 of all large events involve equipment misoperation or human errors among other factors\cite{NERC}. 
This illustrates why stability drawn from the network itself would be desirable.

\section*{The dynamics}
We represent the power grid as a network of power generators and substations (nodes) connected by
power transmission lines (links). 
The nodes may thus consume, produce, and distribute power,  while the links transport power 
and may include passive elements with resistance, capacitance, and inductance.  
The state of the system is determined using power flow calculations 
given the power demand and other properties of the system 
(see Methods),
which is a procedure we implement for the  systems in 
Table~1.
In possession of all variables that describe the steady-state alternating current
flow, we seek to identify the conditions under which the generators can remain stably synchronized. 

The starting point of our analysis is the equation of motion 
\medskip
\begin{equation}\label{eq1}
\frac{2H_i}{\omega_R}\frac{d^2\delta_i}{d t^2}={P_m}_i - {P_e}_i,  
\medskip
\end{equation} 
the so-called 
swing equation (ref.\ \citen{blue} and Methods), which, along with the power flow equations, describes the dynamics of generator $i$.
The parameter $H_i$ is the inertia constant of the generator, 
$\omega_R$ is the reference frequency of the system, 
${P_m}_i$ is the mechanical power provided by the generator, and ${P_e}_i$ is the power demanded of the generator 
by the network (including the power lost to damping). 
In equilibrium, ${P_m}_i={P_e}_i$ and the frequency $\omega_i\equiv\dot{\delta_i}$ remains equal to a common constant for all $i$, which is a sufficient condition for
the frequency synchronization expressed in equation \eqref{eq0}.

When the system changes because of a fault or a fluctuation in power demand, so does ${P_e}_i$.  According to 
equation \eqref{eq1}, immediately after the change the difference between the power demanded and the power supplied by
the generator is 
compensated by either increasing or decreasing the angular momentum from the generator's rotor. 
The change in angular momentum does not remove the difference between ${P_e}_i$ and 
${P_m}_i$; it only compensates for it until the spontaneous or controlled response of ${P_e}_i$ and ${P_m}_i$ 
can adjust the r.h.s.\ of  equation \eqref{eq1}. 
It is through this process that the generators can re-achieve the synchrony they may have
lost in entering the transient state. But what are the properties of the network connecting the generators?

To address this question, we will first represent the power consumed 
at nodes of
the network as equivalent impedances\cite{grey}. 
This involves the assumption that the network structure is fixed and the power demand 
is constant, which is 
appropriate because we are interested in the stability of the system on short time scales. 
For a given 
component,  
the corresponding admittance $Y$ (the reciprocal of the impedance) is related to 
the voltage difference 
$V$
and the complex conjugate of the 
power $P$  through $Y=\bar{P}/|V|^2$.  
Thus,  the admittances are usually complex.

\section*{The network structure}

The admittance matrix {\bf Y$_0$}$=(Y_{0ij})$ is defined analogously to the Laplacian matrix in
oscillator networks: $Y_{0ij}$ is the negative of the admittance between nodes $i$ and  $j\neq i$,
while $Y_{0ii}$ is the sum of all admittances connected to node $i$ (the row sum is nonzero, however, 
due to  finite impedances to the ground).
If \textbf{V} is the vector of node voltages and \textbf{I} is the vector of currents injected into the 
system at the nodes, then, according to the Kirchhoff's law, $\textbf{I}=\textbf{Y$_0$V}$. Because the 
non-generator nodes 
are now regarded as constant impedances, all nodes have zero injection currents except for the 
generator nodes; this can be used to represent the effective 
interactions among the generators through a reduced network. 

Indeed, rearranging \textbf{Y$_0$} such that the first $n$ indices correspond to the generator
nodes and the remaining $r$ indices correspond to non-generator nodes,
one has
\begin{equation}
\left(
\begin{array}{clcr}
\mbox{\bf I$_n$} \\
\mbox{\bf 0}
\end{array}\right)=
\left(
\begin{array}{clcr}
\mbox{\bf Y$_0$}_{n\times n} & \mbox{\bf Y$_0$}_{n\times r} \\
\mbox{\bf Y$_0$}_{r\times n} & \mbox{\bf Y$_0$}_{r\times r}
\end{array} \right)
\left(
\begin{array}{clcr}
\mbox{\bf V$_n$} \\
\mbox{\bf V$_r$}
\end{array} \right).
\end{equation}
The system is then converted to
$\textbf{I}_n=\textbf{Y}\textbf{V}_n$ by eliminating
$\textbf{V}_r$, a procedure known as Kron reduction\cite{blue}, where the resulting {\it effective} admittance matrix is
\begin{equation}
\textbf{Y}=\textbf{{Y$_0$}$_{n\times n}$}-\textbf{{Y$_0$}$_{n\times r}$}
\textbf{{Y$_0$}$_{r\times r}^{-1}$}\textbf{{Y$_0$}$_{r\times n}$}.
\end{equation}
The existence of the matrix \textbf{Y$_0$}$_{r\times r}^{-1}$
follows from the assumed uniqueness of the voltage vector \textbf{V} (with 
respect to a reference voltage). 
In a system with $n$ generators, the symmetric $n\times n$ 
matrix {\bf Y} defines a network in which generator
nodes $i$ and $j\neq i$ are connected by a link of strength 
determined
by the effective admittance $-Y_{ij}$.
The effective admittances are dominantly imaginary (the real part is smaller by up to one order of magnitude 
in the power grids we consider, as shown in 
 Table~2). 
This indicates, as it will become clear below, that capacitances and inductances  
play an important role in the synchronization of generators.

The effective network connecting the generator nodes, represented by \textbf{Y}, 
is significantly different from the physical network of transmission lines represented by \textbf{Y$_0$}, as shown in Fig.~1 for the power grid of Northern Italy.
If the physical network is 
connected, 
as observed for this and any other system that we can treat as a single power grid,  then the effective network will be fully connected\cite{dorfler_ieee2010}, meaning that any two generators are directly coupled through 
an effective admittance.
This property of the effective network holds true under the assumption that the network is in a steady state, and is in contrast to the structure of the corresponding physical network, which is typically 
only sparsely
connected.
The connection strengths are not uniform in both the physical and the effective network (Fig.~1).
However, the distribution of weighted degrees (the sum of the absolute values of all admittances connected to a node) is generally more homogeneous for \textbf{Y} than for \textbf{Y$_0$}, as illustrated in 
Table~1 
for several power networks. This degree homogeneity is a property of interest since it is known that it facilitates synchronization in networks of diffusively coupled oscillators\cite{nish2003,motterSync}. 
Surprisingly, as we show next, in power-grid networks this property can be less determinant for synchronization than is 
 the relation of certain generator parameters to the network structure.

\section*{The stability condition}
To establish an interpretable relationship between synchronization robustness and power-grid parameters, as well as to provide a condition that is necessary for any stricter form of stability, we focus on the short-term stability of the synchronous states under small perturbations.
We thus
linearize  equation \eqref{eq1} around an 
equilibrium (synchronous) state, 
associated with electrical power $P^{*}_{ei}$ and mechanical power $P^{*}_{mi}$
and represented by $\delta^{*}_i$ and $\omega^{*}_i$. 
Assuming that 
$\delta_i=\delta^{*}_i+\delta'_i$, $P_{ei}=P^{*}_{ei}+P'_{ei}$, and 
$P_{mi}=P^{*}_{mi}+P'_{mi}$, the linearized equation reads 
\begin{equation}
\frac{2H_i}{\omega_R}\frac{d^2 \delta'_i}{dt^2}= \frac{\partial P_{mi}}{\partial \omega_i} \omega'_i
                    -\frac{\partial P_{ei}}{\partial \omega_i} \omega'_i
                    -\sum_{j=1}^n \frac{\partial P_{ei}}{\partial \delta_j} \delta'_j,
\label{q:powerCoupOrig}
\end{equation}       
where we neglect the dependence of the mechanical power on changes in the 
phase
${\delta}_i$, 
and denote $\omega'_i=\dot{\delta}'_i$. For the first 
term on the right-hand side, we assume the droop equation $\partial P_{mi}/\partial \omega_i=-1/(\omega_R R_i)$, where $R_i>0$ 
is a regulation parameter. For the second term, we assume there is a constant damping
coefficient $D_i>0$ such that $\partial P_{ei}/\partial \omega_i=D_i/\omega_R$. 
The third term can be 
obtained from
$P'_{ei}=
D_i \omega'_i/\omega_R +
\sum_{j=1}^{n} E_i E_j (B_{ij}\cos\delta^*_{ij} - G_{ij}\sin\delta^*_{ij})\delta'_{ij}$,
where  $\delta^*_{ij}=\delta^*_i-\delta^*_j$, $\delta'_{ij} = \delta'_i-\delta'_j$,
$G_{ij}$ and $B_{ij}$ are the real and imaginary components of $Y_{ij}$, and $E_i$ is the internal-voltage magnitude of
the $i$th generator\cite{grey}. 
We denote the $n$-dimensional vectors of $\delta'_i$ and $\dot{\delta}'_i$
by
\textbf{X$_1$} and \textbf{X$_2$}, respectively.

This
yields a coupled set of $2n$ first-order equations,
\begin{eqnarray}\label{x1}
\dot{\textbf{X}}_1 &=& ~~\textbf{X$_2$},\\ 
\dot{\textbf{X}}_2 &=& -\textbf{P}\textbf{X}_1 - \textbf{B}\textbf{X}_2, \label{x2}
\end{eqnarray}
where $\textbf{P}=(P_{ij})$ is the zero row sum matrix given by
\begin{eqnarray}\label{eqn:P}
P_{ij}= \left\{
\begin{array}{cl}
\frac{\omega_R E_i E_j}{2H_i} (G_{ij}\sin\delta^*_{ij} - B_{ij}\cos\delta^*_{ij}), & i\neq j,\\
       -\sum_{k\ne i} P_{ik},  & i=j, 
\end{array} 
       \right.       
\end{eqnarray}
and $\textbf{B}$ is the diagonal matrix of elements $\beta_i=(D_i+1/R_i)/2H_i$. 
If we now assume this parameter $\beta_i$ to be the same for all generator nodes, 
equations \eqref{x1} and \eqref{x2}
can be diagonalized using the substitution $\textbf{Z}_1=\textbf{Q}^{-1}\textbf{X}_1$ 
and $\textbf{Z}_2=\textbf{Q}^{-1}\textbf{X}_2$, where \textbf{Q} is a matrix of eigenvectors 
of \textbf{P} so that \textbf{J} = \textbf{Q$^{-1}$PQ} is the diagonal matrix of the corresponding
eigenvalues.
This leads to $\dot{\textbf{Z}}_1=\textbf{Z}_2$ and 
$\dot{\textbf{Z}}_2=-\textbf{J}\textbf{Z}_1-\beta\textbf{Z}_2$, where $\beta$ is the common value of the parameter $\beta_i$.
Naturally,
$\beta_i$ is not the same for all generators in the power grids we consider, as evidenced by the normalized
standard deviation in  
Table~1 
(column 7),  
but this auxiliary assumption will allow us
to derive results to enhance synchronization stability of power grids with nonidentical $\beta_i$.

Equations
\eqref{x1} and \eqref{x2} are then reduced to $n$ decoupled $2$-dimensional
systems of the form
\begin{eqnarray}\label{zeta}
\;\;\;\;\; \dot{\zeta}_j=
\left(
\begin{array}{cc}
 0  &  1 \\
-\alpha_j & -\beta
\end{array}
\right) 
\zeta_j,\;\;\;
\zeta_j\equiv
\left(
\begin{array}{c}
Z_{1j}\\
Z_{2j}
\end{array}
\right), 
\end{eqnarray}
where $\alpha_j$ is the $j$th eigenvalue of the coupling matrix \textbf{P}. 
The matrix \textbf{P} always  has a null eigenvalue associated with a uniform 
shift of all phases, 
determined
by $\textbf{X}_1 \propto (1,\dots,1)^T$, 
which corresponds to two perturbation eigenmodes (one for $\textbf{X}_2=0$ and one for $\textbf{X}_2= - \beta\textbf{X}_1$)
that do not affect the synchronization condition in equation \eqref{eq0}.  
We denote this eigenvalue by $\alpha_1$ and exclude the corresponding eigenmodes
from subsequent analysis.

The stability of the synchronous state is governed by the Lyapunov exponents
of system \eqref{zeta},
\begin{equation}\label{lyap}
\lambda_{j_\pm}(\alpha_j,\beta)= -\frac{\beta}{2} \pm \frac{1}{2}\sqrt{\beta^2 - 4\alpha_j}\, .
\end{equation}
The synchronous state is stable iff the real component of every Lyapunov exponent 
is negative for  all $j\ge 2$. That is, the stability condition translates to
\begin{equation}
\label{eq:max_eq}
\max_{\{\pm\}} Re{\lambda_{j_\pm}} \leq 0, \;\;\; \mbox{for} \; j=2,\ldots, n. 
\end{equation}
Because equation \eqref{zeta} is formally identical for all $j$, it is useful to drop the index $j$ in equation \eqref{lyap}
and define the function $\Lambda_{\beta}(\alpha)=\max_{\{\pm\}}{ Re{\lambda_{\pm}}(\alpha,\beta)}$
for fixed  $\beta$ and tunable parameter $\alpha$. 
This function can be interpreted as a {\it master} stability function: the stability of the system is determined by whether all the eigenvalues $\alpha_2,\dots,\alpha_n$ fall within the negative region of 
$\Lambda_{\beta}$.  This stability condition is of the form of those in refs.\ \citen{pecora_prl_1998} and \citen{PhysRevE.61.5080} for coupled oscillators, as explicitly shown in Methods.

The matrix \textbf{P} is generally asymmetric, meaning that the influence of a generator on another generator is
not necessarily identical to the influence of the second on the first. But we argue that the eigenvalues of \textbf{P}
are all real over a range of conditions. We first factor the matrix as  
$\textbf{P}= \textbf{H}^{-1}\textbf{P}'$, where $\textbf{H}$ is the 
diagonal matrix of inertia constants $H_i$, to then note (inspired by a similar argument in ref.\ \citen{motterSync})
that the eigenvalues of $\textbf{P}$ equal the eigenvalues of the partially symmetrized 
matrix $\textbf{P}''= \textbf{H}^{-1/2}\textbf{P}'\textbf{H}^{-1/2}$. 
The matrix
$\textbf{P}''$
is a zero row sum matrix of off-diagonal elements 
$\frac{\omega_R E_i E_j}{2\sqrt{H_iH_j}} (G_{ij}\sin\delta^*_{ij} - B_{ij}\cos\delta^*_{ij})$.
We now observe that,  due to the small real-to-imaginary component ratio of the admittances 
and the predominance of small phase differences $\delta^*_{ij}$  
(Table~2), 
the antisymmetric part of $\textbf{P}''$ is usually much 
smaller than the symmetric one; the 2-norm is over one order of magnitude 
smaller for the systems considered here, as shown in 
Table~3. 
If the eigenvalues of the symmetric part of $\textbf{P}''$, which are 
necessarily real, 
are nondegenerate (as generally observed in practice), 
then we can show that the eigenvalues remain strictly real in the 
presence of a small antisymmetric component.
Therefore, while our stability condition can be used in general, in what
follows we will assume that \textbf{P} has real eigenvalue spectra, 
a condition observed to be satisfied in all of our simulations. 

For real $\alpha$,  the stability region defined by $\Lambda_{\beta}(\alpha)<0$ corresponds to $\alpha>0$.
Indeed, the function $\Lambda_{\beta}$ is zero for $\alpha = 0$ and becomes
positive and increasingly larger for smaller $\alpha<0$; in the interval $0<\alpha\le \beta^2/4$, 
the function $\Lambda_{\beta}$ is negative and decreases as $\alpha$ increases. 
For $\alpha>\beta^2/4$,  the function $\Lambda_{\beta}$ is negative and remains
constant since the radicand in equation \eqref{lyap} is negative, rendering 
$\lambda_{j_\pm}$ 
to be complex. This behavior is summarized in Fig.~2a for different 
values of $\beta$, which further illustrates that
varying $\beta$ changes the behavior of the Lyapunov exponents as a function of $\alpha$
and, in particular, $\Lambda_{\beta}(\alpha\ge \beta^2/4) = -\beta/2$, but the stability 
condition that $\alpha_j>0$ for all $j\ge 2$ does not change.
Moreover, for a given $\beta>0$, the master stability condition does not depend on
whether this factor $\beta$ 
 is due to the presence of the regulation parameter $R_i$, the presence
of damping $D_i$, or a combination of both.  However, all other parameters left constant,
both $\beta$ and the eigenvalues $\alpha_j$ decrease as the inertia parameters $H_i$ are
uniformly increased. This has the effect of reducing 
$|Re{\lambda_{j_\pm}}|$
 and hence the
strength of the linear stability or instability.

\section*{Enhancement of synchronization stability}

We now apply this formalism to enhance the stability of the synchronous states by tuning the parameters of the generators.  For $\beta_i = \beta$, the synchronization stability is solely determined by the master stability function $\Lambda_\beta(\alpha)$ at $\alpha = \alpha_2$, the smallest nonzero (real) eigenvalue of {\bf P}. This follows from the non-increasing dependence of $\Lambda_\beta(\alpha)$ on $\alpha$.  For $\alpha_2>0$, which corresponds to stable synchronization, the quantity $\Lambda_\beta(\alpha_2)$ as a function of $\beta$ attains its minimum at 
\begin{equation}
\beta = \beta_\text{opt} \equiv 2\sqrt{\alpha_2}
\label{new_equation}
\end{equation}
 (see Fig.~2b). 
Given a network structure and a steady state, and hence a fixed value of $\alpha_2$, this point of maximum stability can be achieved by adjusting one of the generator parameters. 

Specifically, one can ensure that $\beta_i = \beta_\text{opt}$ for all generators by adjusting the droop parameter $R_i$ to
\begin{equation}\label{eqn:opt_R}
R_i = \frac{1}{4H_i\sqrt{\alpha_2} - D_i}, \quad i = 1,\ldots,n,  
\end{equation}
or the damping coefficient $D_i$ to
\begin{equation}\label{eqn:opt_D}
D_i = 4H_i\sqrt{\alpha_2}  -\frac{1}{R_i}, \quad i = 1,\ldots,n,  
\end{equation}
while keeping the other parameters unchanged.
The tuning of the former, $R_i$, is most suitable for off-line optimization of stability, since the time scales associated with this parameter\cite{grey} are usually larger than those associated with 
typical instabilities.   
The latter, $D_i$, can be adjusted dynamically at very short 
time scales\cite{grey}
 and hence is relevant for online optimization and fine-tuning under varying operating conditions that affect $\alpha_2$.
Since equation~\eqref{new_equation} can be satisfied exactly, 
this approach leads to
the fastest possible asymptotic convergence rate to the synchronous state among all systems with the same network structure and identical $\beta_i$ values. 

More important,
 even when the $\beta_i$ values are not identical, as in the case of real power grids, adjusting  $R_i$ according to equation \eqref{eqn:opt_R} or $D_i$  according to equation \eqref{eqn:opt_D} in general results in substantially improved stability. For example, consider the three test systems and the Guatemala power grid in 
Table~1, 
which are all stable before any parameter adjustment.  
As these systems have nonidentical $\beta_i$ values 
(Table~1, column 7), 
their stability was determined directly from equations~\eqref{x1} and \eqref{x2} before diagonalization; 
the stability is
quantified by $\lambda^{\max}$, the largest Lyapunov exponent excluding the null exponent associated with the uniform shift of phases.
A naive approach based on parameter adjustment that homogenizes  $\beta_i$ to their average $\bar\beta = \sum_i \beta_i/n$ does not necessarily 
improve stability, as illustrated by the opposite effect 
observed in the Guatemala system 
(Table~1, columns 8 and 10). 
However, the adjustment  to $\beta_\text{opt}$ improves 
stability in all systems, by a factor ranging from 5 to nearly 90 in terms of $\lambda^{\max}$
(Table~1, column 11). 

This demonstrates that the mechanism by which synchronization stability can be enhanced requires not only homogenization of the function $\beta_i$ of the generator parameters $H_i$, $D_i$, and $R_i$, but also matching of this function with the structural and the dynamical state of the network represented in $\alpha_2$. 
Furthermore, 
we can show that the maximum Lyapunov exponent $\lambda^{\max}$ is locally minimum at $\beta_i = \beta_\text{opt}$ along any given direction in the $\beta_i$-space for all systems we 
consider,
which is a necessary condition for this parameter assignment to be locally optimal for synchronization stability.  The approach is therefore appropriate
for enhancing the stability of 
power networks for which the synchronous state is already stable. 
If the synchronous state is 
not stable,
as in the case of the Northern Italy and Poland systems for the dynamic parameters we consider, we can first make it stable by adjusting the generators' transient reactances $x'_{d,i}$ (see Methods and Table~1, column 9).
Then, the synchronization stability can be further enhanced by the adjustment of $\beta_i$ to $\beta_\text{opt}$, as illustrated in Fig.~\ref{fig3} for the Northern Italy system.
Taken together, these results provide a systematic 
approach
to strengthen stability in power-grid networks.

\section*{Discussion}

Power grids are dynamic entities whose structure is history-dependent and evolves in  a decentralized way that is often determined by market. 
The consequent difficulty to  rationally design and modify the network may be seen as a major limiting factor in optimizing synchronization in these systems. 
Previous
theoretical 
studies of synchronization
in oscillator networks
have shown that the structure of the interaction network is a determinant factor for the dynamical units to synchronize\cite{review2}. In the most studied case of diffusively-coupled oscillators, it has been further shown that certain network structures that inhibit synchronization
can in fact facilitate synchronization when in presence of other synchronization-inhibiting network structures, such as negative interactions\cite{nishikawa2010}. 
But in power grids, the relevant network  
is not simply the physical network of transmission lines, and this causes 
the structure and dynamics to be more intimately related than in such idealized models (see Methods). 
In part because of this,  here we have been able to show that the stability of synchronous states in power grids can be enhanced by 
tuning
parameters of the dynamical units rather than the network.

Such enhancement can be implemented through the condition expressed by $\beta_i = 2 \sqrt{\alpha_2}$.
The r.h.s.\ of this equation accounts for the network structure and indirectly depends on the generators (see Methods) while the l.h.s.\ depends only on the generator parameters $H_i$, $R_i$, and $D_i$, among which the latter two do not appear on the r.h.s.  
Of these,  
the droop parameter $R_i$ offers a solution that accounts for slow changes in demand and the long-term evolution of the network. 
We argue that the other parameter, the damping coefficient $D_i$, has the flexibility required to cope with rapid changes caused by a fault or fluctuation, since the adjustment of this parameter can be realized through very fast control loops (e.g., by adding power system stabilizers).
Our optimization scheme can be effective even when modeling load dynamics\cite{Bergen:1981kx} 
(Fig.\ \ref{fig:away_bif}), 
and it is complementary to an approach proposed recently\cite{Mallada:2011lr} to mitigate saddle-node instability by adjusting power scheduling or line impedances 
(Fig.\ \ref{fig:compare}).
Conversely, the adjustment of power scheduling and/or line impedances can also
be effective within 
the reduced network model used in our study
(Fig.\ \ref{fig:MT_in_classical}).

We suggest that 
our
findings are potentially important for ongoing research on smart grids, which is making it ever more important to understand optimization and control of power-grid dynamics. While the study of stabilization via damping of oscillations has a long history\cite{gooi1981,dobson}, the proposed coordinated 
 tuning of generators to enhance stability is a timely approach in view of the upcoming availability of system-wide data from phasor measurement units\cite{pmu2007}. These data will provide accurate information about the synchronization state of 
 the power grid, which 
can be integrated with the online control of generator parameters. Therefore, besides contributing to the devise of more robust systems, our findings provide insights for the development of efficient controllers. Such controllers may help advance research on self-healing systems that can recover from failures in real time. 
This
can lead to systems that are less prone to cascading failures, which, as amply discussed in the literature on 
interdependent networks\cite{Rinaldi:2001fk}, 
have consequences that transcend the power grid itself.

Our analysis of power-grid dynamics also provides a fresh view of synchronization in complex networks in general.  By identifying as the leading factor for stability the relation between the specifics of the dynamical units and the network structure, it contrasts with most previous studies, which focused on the role of the 
network structure alone (analyses of correlations between the natural frequencies of phase oscillators and their connection topology are among the few exceptions\cite{Carareto:2009fk,PhysRevE.80.066120,kelly:025110,Markus:2008uq,PhysRevLett.106.128701}).
Aside from power grids and other technological applications, this is likely to have implications for natural systems, such as biological ones, where the dynamical units co-evolve with the network structure\cite{Garlaschelli:2007fk}.  Moreover, 
our analysis establishes 
a master stability condition for the synchronization of power generators, casting the problem in the framework of oscillator networks\cite{pecora_prl_1998} used to investigate collective behavior and nonlinear phenomena in a wide range of complex systems. 
This facilitates comparison across different domains, and, we hope, will inspire similar developments on  the tailoring of collective behavior 
in
other real systems.

\section*{Methods}

\paragraph{Power flow calculations and synchronous states.}
The power and voltage at each node were determined given  
the net injected real power 
and the voltage magnitude at each generator node, the power demand at each non-generator node, the reference voltage 
phase, 
the admittance of each power line, and the capacities of the network components.  These power flow calculations were performed using the Power Systems Analysis Toolbox (PSAT)\cite{psat}.
To identify the synchronous states, 
the phase
$\delta^*_i$ of generator $i$ was determined from the power and voltage at node $i$ assuming the classical model\cite{grey}. In this model, the generator is represented by a voltage source with constant magnitude $E_i$ and variable 
phase
$\delta_i$ that is connected to the rest of the network through a transient reactance $x'_{d,i}$ and a terminal. Throughout the paper, we represent each generator and its terminal by a single node,  except in the Kron reduction, where the terminals are treated as independent nodes to be eliminated.

\paragraph{Derivation of the swing equation.}
Equation~\eqref{eq1} 
can be derived by setting the rate of change of the angular momentum of the rotor equal to the net torque acting on the rotor:
\begin{equation}
J \frac{d^2\delta_i}{d t^2}={T_m}_i - {T_e}_i, 
\end{equation}
where $J$ is the moment of inertia in kg$\cdot$m$^2$, ${T_m}_i$ is the mechanical torque in N$\cdot$m accelerating the rotor, and ${T_e}_i$ is the typically decelerating torque in N$\cdot$m due to electrical load in the network.
Multiplying both sides by $\omega_i$ and using the fact that the torque in N$\cdot$m multiplied by the angular velocity in radians per second gives the power in watts, the equation can be written in terms of power:
\begin{equation}
J\omega_i \frac{d^2\delta_i}{d t^2}={P_m}_i - {P_e}_i.
\end{equation}
To make ${P_m}_i$ and ${P_e}_i$ per unit quantities, we divide both sides of the equation by the rated power $P_R$ (used as a reference).
The factor $J\omega_i$ then becomes $2H_i/\omega_i$, where we defined the inertia constant $H_i = W_i/P_R$ (in seconds) and the kinetic energy of the rotor $W_i = \frac{1}{2}J\omega_i^2$ (in joules).
Noting that $\omega_i$ is approximately equal to the reference frequency $\omega_R$ in systems close to synchronization, we obtain 
equation~\eqref{eq1}.
A more detailed description of this derivation can be found in 
ref.~\citen{grey} 
(second edition, pp.\ 13--16).

\paragraph{Relation to the master stability formalism.}
Equations~\eqref{x1} and \eqref{x2} 
when written individually for each node $i$, can be expressed as
\begin{equation}\label{eqs67i}
\frac{d}{dt}\begin{bmatrix}\delta'_i\\ \dot{\delta}'_i \end{bmatrix}
= \begin{bmatrix} 0 & 1\\ 0 & -\beta_i \end{bmatrix}\begin{bmatrix}\delta'_i\\ \dot{\delta}'_i \end{bmatrix}
 + \sum_{j=1}^n P_{ij}\begin{bmatrix} 0 & 0\\ -1 & 0 \end{bmatrix}
 \begin{bmatrix}\delta'_j\\ \dot{\delta}'_j \end{bmatrix}.
\end{equation}
This is in the same general form as the variational equation for the class of coupled oscillators considered in 
ref.~\citen{pecora_prl_1998}, 
which for a synchronous state $\textbf{s} = \textbf{s}(t)$ is
\begin{equation}\label{vareq_ref31}
\frac{d}{dt}\textbf{x}_i = D\textbf{F}(\textbf{s})\cdot\textbf{x}_i + \sigma\sum_{j}G_{ij}D\textbf{H}(\textbf{s})\cdot\textbf{x}_j,
\end{equation}
where $\dot{\textbf{x}}_i = \textbf{F}(\textbf{x}_i)$ describes the node dynamics, $\textbf{H}$ is the coupling function, and $\sigma G_{ij}$ represents the strength of coupling from node $j$ to $i$.
The factors $D\textbf{F}(\textbf{s})$ and $D\textbf{H}(\textbf{s})$ in equation~\eqref{vareq_ref31} are both constant matrices in equation~\eqref{eqs67i}, and $P_{ij}$ in equation~\eqref{eqs67i} corresponds to $\sigma G_{ij}$ in equation~\eqref{vareq_ref31}, which are relations that can be used to derive $\Lambda_\beta(\alpha)$ from 
equations~\eqref{x1} and~\eqref{x2} 
based on 
the results in 
ref.\ \citen{pecora_prl_1998}. 
Therefore, while the formalism in 
ref.\ \citen{pecora_prl_1998} 
cannot be directly applied to 
equation~\eqref{eq1}, it can be applied near the synchronization manifold, which is a procedure previously proposed for a broad class of coupled 
oscillators\cite{PhysRevE.61.5080}.

\paragraph{Adjusting transient reactances for stability.}
We first observe that for $B_{ij}>0$, which is the most frequent case (Table~4), the cosine term in equation \eqref{eqn:P} is a destabilizing factor if $|\delta^*_{ij}| > \pi/2$. This instability often arises when a few specific generators have phases that are significantly different from those of the other generators in the synchronous state, which can be due to these generator's transient reactance $x'_{d,i}$. By identifying these generators through the negative sign of the corresponding diagonal elements of {\bf P} and reducing their $x'_{d,i}$, the instability can be suppressed.
Applying this 
procedure
 to the Northern Italy and Poland networks indeed turns the systems stable 
(Table~1, column 9).

\paragraph{Relation between structure and dynamics.}
The interactions between generators
in a power grid are in principle determined by the effective admittances, 
which
are dominantly imaginary 
(Table~2).
Thus, these interactions too have a sign; an inductive reactance ($B_{ij}>0$) suggests a positive interaction, and a capacitive reactance ($B_{ij}<0$) a negative one.
However,
it can be seen from equation~\eqref{eqn:P} that the interactions between the generators are in fact determined by $B_{ij}\cos\delta^*_{ij}$.
In order for $\alpha_2>0$, as required for stable synchronization, the terms $B_{ij}\cos\delta^*_{ij}$ have to be dominantly positive, as observed in real systems
(Table~4).
One can make any of these terms positive by 
having $B_{ij}$ and $\cos\delta^*_{ij}$
both positive 
or both
negative, where having both factors positive is far more common in real systems than both negative (Table~4). 
While
$B_{ij}$ is mainly structural given the power flow solution, the parameter $\delta^*_{ij}$ can be adjusted by changing specific properties of the generators, 
such as their transient reactances. 
Therefore, the interactions 
are in reality determined not only by the network of effective admittances, but also by the system-level (alternate current) dynamics, which is in turn molded by the properties of the dynamical units. As a result, important aspects in the network structure can be emulated by modifying tunable parameters of the dynamical units.

\paragraph{Power-grid data.}
The data required for power flow calculations were obtained as follows.  For the  10-generator system, known as the New England test system, the parameters were taken from ref.\ \citen{pai1989energy}. 
For the 3- and 50-generator systems, the parameters were taken from 
refs.\ \citen{grey} 
and \citen{141684}, respectively.  The data for the Guatemala and Northern Italy systems were provided by F.\ Milano (University of Castilla -- La Mancha), 
and the data for the Poland system were provided as part of the MATPOWER software package\cite{matpower}.  
The 
dynamic
data required to calculate the synchronous state, determine its stability, and simulate 
equation~\eqref{eq1}, are not all available for the real power grids, and 
were obtained as follows. For all systems, we assumed that before any optimization the damping coefficient and droop parameter satisfy $D_i+1/R_i=50$ per unit  for all generators. The parameters $x'_{d,i}$ and $H_i$ are available for the three test systems from the respective references mentioned above.  For the Guatemala, Northern Italy, and Poland systems, we estimated $x'_{d,i}$ and $H_i$ using the strong correlation observed  
in the test systems between each of these parameters and  the power $P_i$  injected by generator $i$ into the network, as shown in Fig.\ \ref{fig:dyn_data}.
The estimated values are
$x'_{d,i} \approx 92.8P_i^{-1.3}$  and $ H_i \approx 0.04 P_i$, where $P_i$ is in megawatts, $x'_{d,i}$ is  in per unit, and $H_i$ is in seconds.


\begin{addendum}  \baselineskip23pt
\item The authors thank F.\ Milano for providing power-grid data, E.\ Mallada for sharing unpublished simulation details, and F. D\"{o}rfler for insightful discussions.
This work was supported by NSF under Grants DMS-1057128 and DMS-0709212, the
LANL  LDRD project Optimization and Control Theory for Smart Grids, and a Northwestern-Argonne Early Career Investigator Award for Energy Research to A.E.M.
\item[Author contributions] 
All authors 
contributed to the design of the research and analytical
calculations. 
S.A.M., M.A., and T.N. 
performed the numerical simulations.  A.E.M.
and T.N. wrote the paper, and A.E.M supervised the project. 
\end{addendum}

\clearpage

\begin{figure}[tbp]
\bfseries
\begin{center}
\includegraphics[width=.7\textwidth]{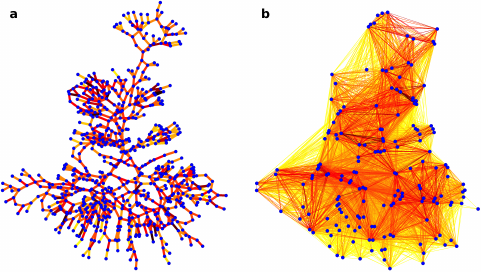}
\caption{
\baselineskip15pt
\textbf{Physical versus effective network for the power grid of Northern Italy. }\mdseries
\textbf{a}, Representation of the 
physical network of transmission lines, which has 678 nodes and 822 links.  
\textbf{b},~Representation of the network of effective admittances, which is an all-to-all
network with 170 nodes corresponding to the generators in the system (a subset of the nodes in panel \textbf{a}). 
The color scale of the lines indicates the link weights, 
ranging from yellow to red to black (scaled differently for each panel), 
defined as the absolute value of the corresponding admittance.
For clarity, in panel~\textbf{b} we show only the top 50\% highest-weight links.}
\end{center}
\end{figure}
\baselineskip18pt

\begin{figure}[tbp]
\bfseries
\begin{center}
\includegraphics[width=.8\textwidth]{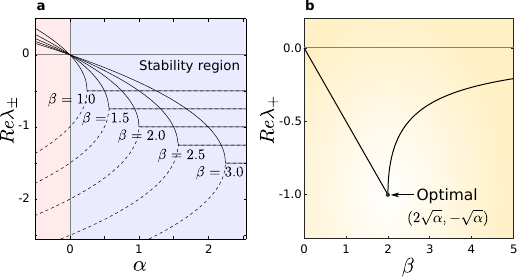}
\caption{
\baselineskip15pt
\textbf{Stability of synchronous states for systems with $\boldsymbol{\beta_i = \beta}$.}\mdseries
\,\,\textbf{a}, Real part of the Lyapunov exponents $\lambda_{+}$ (continuous lines) and $\lambda_{-}$ (dashed lines) 
as functions of $\alpha$ for increasing values of $\beta$ in the range $1.0$--$3.0$. 
Although changing $\beta$ causes 
changes in the shapes of $Re \lambda_{\pm}$, the region of stability defined by $\Lambda_{\beta}(\alpha)<0$ is 
always $\alpha > 0$. 
\textbf{b}, Real part of the Lyapunov exponent $\lambda_{+}$ as a function of $\beta$, which has a minimum at $\beta = 2\sqrt{\alpha}$ (illustrated here for $\alpha = 1$). The stability of synchronous states is maximum at this point.}
\end{center}
\end{figure}

\begin{figure}[tbp]
\bfseries
\begin{center}
\includegraphics[width=\textwidth]{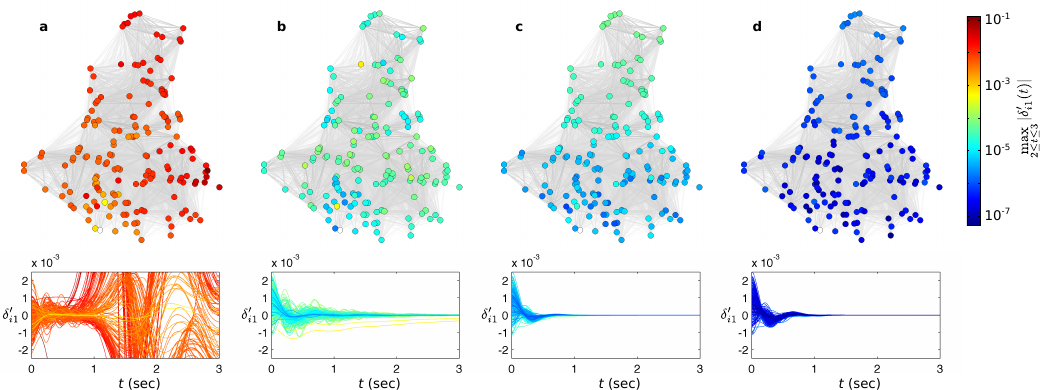}
\caption{
\baselineskip15pt
\textbf{Enhancement of the stability of synchronous states.} 
\mdseries
\,\textbf{a}--\textbf{d}, Response of the original network (\textbf{a}), the network with adjusted $x'_{d,i}$ (\textbf{b}), the network with $\beta_i = \bar\beta$ (\textbf{c}), and the network with $\beta_i = \beta_\text{opt}$ (\textbf{d}) to 
a perturbation
in
the Northern Italy power grid. 
The perturbation 
was
applied to the phase of each generator in the synchronous state  at $t = 0$, and was drawn from the Gaussian distribution with mean zero and standard deviation  0.01 rad.
In each case, the network layout is the same as in Fig.~1b and the bottom panel shows the time evolution of $\delta'_{i1}=\delta_{i1}-\delta^*_{i1}$,  where $\delta_{i1}$ is 
the phase 
of generator $i$  relative to generator $1$ and $\delta^*_{i1}$ is the corresponding 
phase
in the synchronous state. Generator 1, shown as a white node in the network, is used as a reference to 
discount phase drifts common to all generators.
The other nodes and their time-evolution curves are color-coded by the maximum value of $|\delta'_{i1}|$ for $2 \le t \le 3$.  
By adjusting the transient reactance of the generators, the divergence from the unstable steady state is converted to exponential convergence (\textbf{a} and \textbf{b}).  This stability is improved upon adjusting the generator parameters to ensure a common value for $\beta_i$, but is further improved when this common value is tuned to $\beta_\text{opt} = 2\sqrt{\alpha_2}$ (\textbf{c} and \textbf{d}).}
\label{fig3}
\end{center}
\end{figure}

\begin{figure*}
\bfseries
\centering
\includegraphics[width=0.6\textwidth]{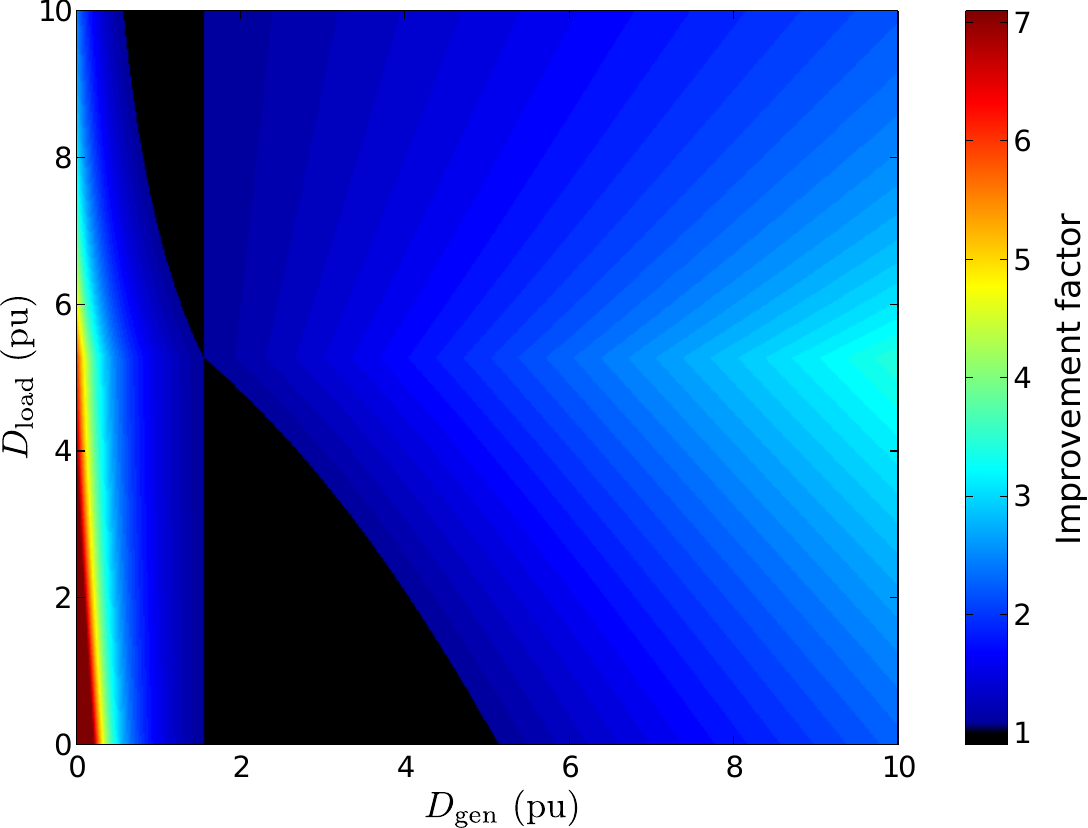}
\caption{\label{fig:away_bif}
\baselineskip15pt
\textbf{Enhancement of synchronization stability in power-grid model incorporating load dynamics.}
\mdseries
Colors indicate the factor by which the stability improves after adjusting all $\beta_i$ to the value $\beta_\text{opt}$, where the darkest red corresponds to any factor $> 7$ and black to any factor $<1$.
The stability of a synchronous state is measured by the largest nonzero Lyapunov exponent $\lambda^{\max}$ computed within the Bergen-Hill model\cite{Bergen:1981kx}.
In this model, instead of eliminating non-generator nodes by the Kron reduction, load dynamics is modeled by assuming that the real power is a linear function of the voltage frequency for all power-consuming nodes, while the voltage magnitude is assumed to be constant for all nodes. 
For illustration, we use the 3-generator system studied in ref.\ \citen{Mallada:2011lr}, with 
the power injection modified to $P_i = 5, 6, 7$ per unit. 
The improvement factor is shown as a function of the parameters $D_\text{gen}$ and $D_\text{load}$, 
where $D_\text{gen}$ is a coefficient given by $(D_i + 1/R_i)/\omega_R$ (assumed be the same for all generator nodes) 
and $D_\text{load}$ is the frequency coefficient (assumed to be the same for all the other nodes).
In most cases, our method yields significant stability improvement, demonstrating its effectiveness beyond the reduced network model.
} 
\end{figure*}

\begin{figure*}[ht]
\bfseries
\centering
\includegraphics[width=0.6\textwidth]{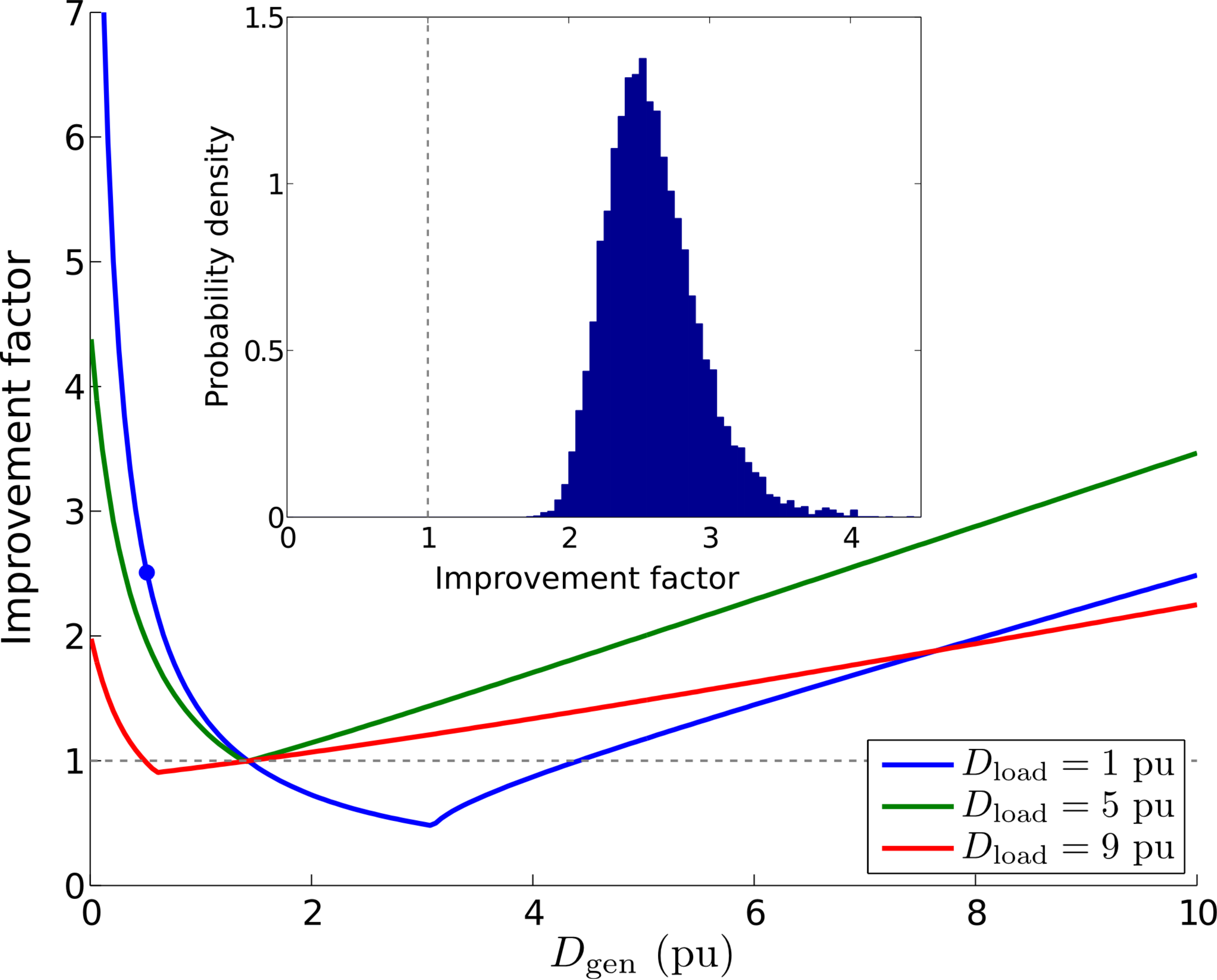}
\caption{\label{fig:compare}
\baselineskip15pt
\textbf{Combination of complementary approaches to mitigate instabilities associated with saddle-node bifurcations.}
\mdseries
As an example, we use the same 3-generator system described in the caption of Fig.\ \ref{fig:away_bif} with   power injection $P_i = 7.994$, $3.006$, $7.000$, simulated using the Bergen-Hill model\cite{Bergen:1981kx}.
For $D_\text{gen} = D_\text{load} = 1$ per unit, the system is near a saddle-node bifurcation, and thus the Lyapunov exponent $\lambda^{\max}$ is negative real and close to zero.
At each iteration of the gradient descent-like method described in ref.\ \citen{Mallada:2011lr}, which changes the power injected by the generators (independently of the values of $\beta_i$ and thus of $D_i$), we compute $\lambda^{\max}$ both with and without adjusting $\beta_i$ to $\beta_\text{opt}$ by tuning the damping coefficients $D_i$.
The improvement factor, measured by 
the ratio between the smallest $\lambda^{\max}$ obtained through the iterative process with and without the $\beta_i$-adjustment, 
is shown as a function of $D_\text{gen}$ for different values of $D_\text{load}$.
In most cases, significant additional improvement results from the adjustment of $\beta_i$, illustrating that near instabilities the two methods can be combined to achieve enhancement not possible by either method alone.
This result is robust against heterogeneity in the network parameters, as illustrated in the inset histogram for the system with $D_\text{gen}=0.5$ and $D_\text{load}=1$ (blue dot in the main plot), where each of 
these
coefficients 
is
independently perturbed according to the Gaussian distribution with mean zero and standard deviation $0.1$.
}
\end{figure*}

\begin{figure*}[ht]
\bfseries
\centering
\includegraphics[width=0.6\textwidth]{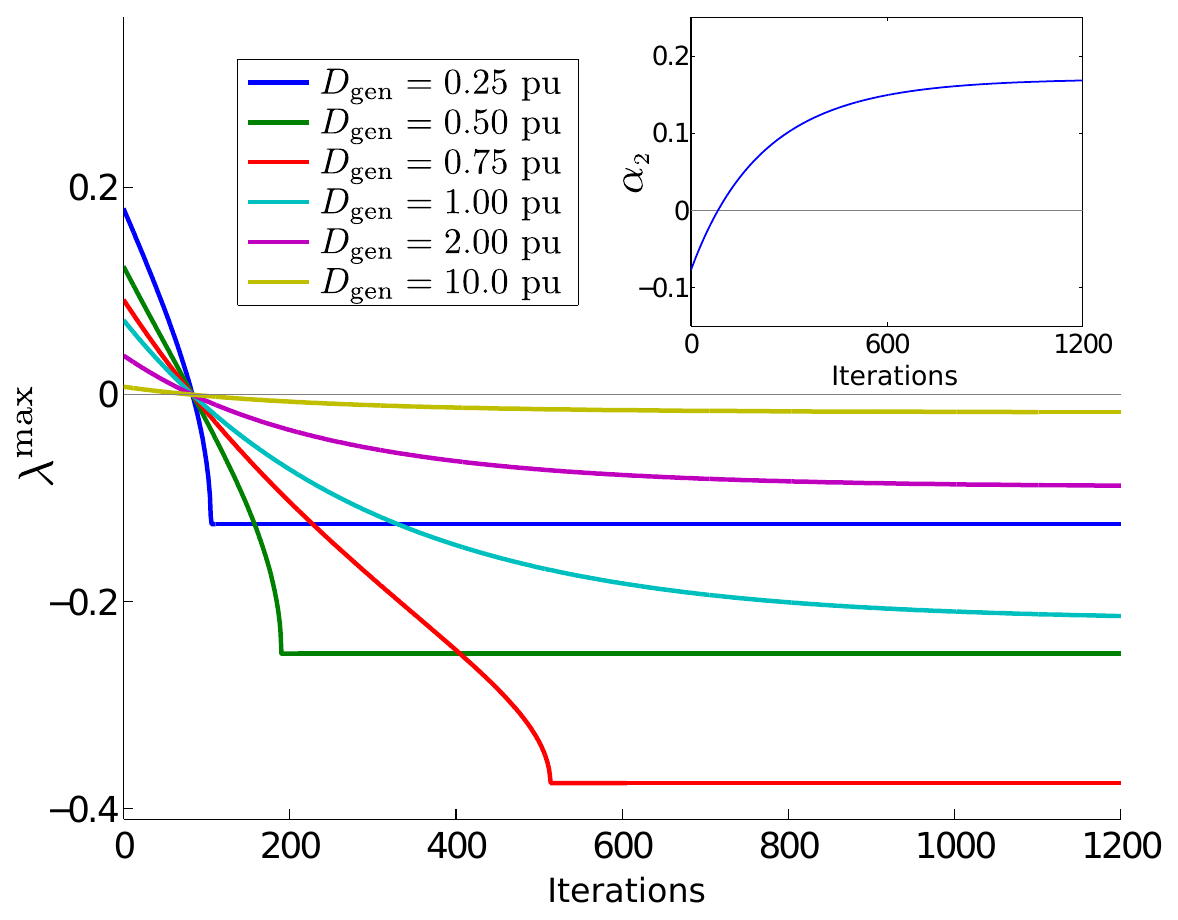}
\caption{\label{fig:MT_in_classical}
\baselineskip15pt
\textbf{Enhancement of synchronization stability 
via tuning of power injection within the reduced network model.
}
\mdseries
We use the same 3-generator system as in Figs.\ \ref{fig:away_bif} and \ref{fig:compare}, where 
all system parameters are the same as in ref.\ \citen{Mallada:2011lr}, except for 
the power injection and consumption, which are initially set to 60\% of the values used in that reference. 
This makes the system unstable, as indicated by the positivity of the Lyapunov exponent $\lambda^{\max}$ for the synchronous state.
The power injections were
then 
adjusted iteratively according to the gradient descent-like scheme of ref.\ \citen{Mallada:2011lr} and $\lambda^{\max}$ is plotted against the number of iterations for several values of the 
coefficient $D_\text{gen}$.
In all cases, the power adjustment stabilizes the synchronous state, which is a consequence of increasing $\alpha_2$ across zero, as shown in the inset. 
Note that $\alpha_2$ is independent of $D_\text{gen}$ because the synchronous state in the 
reduced network
model is unaffected by changes in $D_\text{gen}$.
} 
\end{figure*}

\begin{figure*}[th]
\bfseries
\begin{center}
\includegraphics[width=0.9\textwidth]{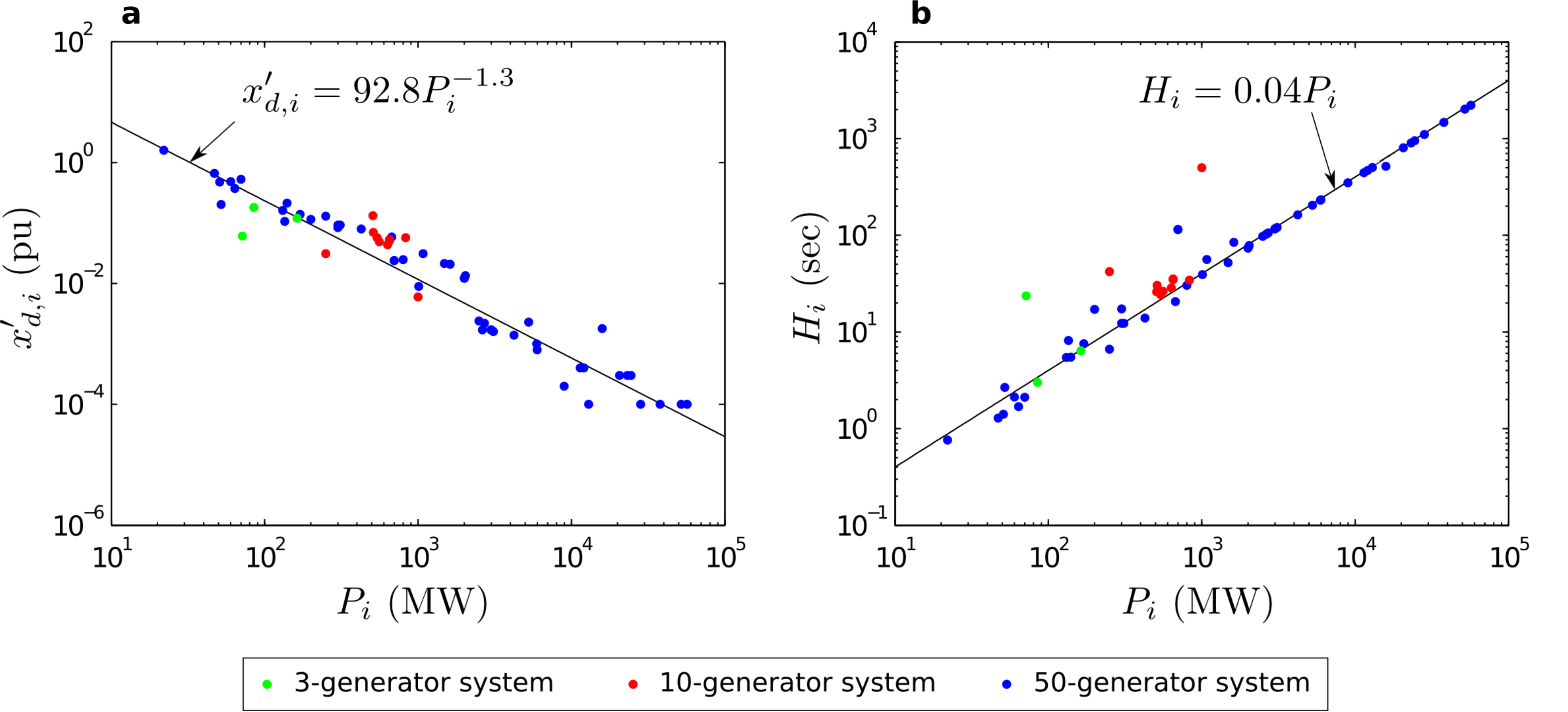} 
\caption{\label{fig:dyn_data}
\baselineskip15pt
\textbf{Relationship between the power produced by individual generators and their parameters in the three test systems of Table~1.} 
\mdseries
\textbf{a}, The transient reactance $x'_{d,i}$ versus the power $P_i$  injected into the network. \textbf{b}, The inertia constant $H_i$ versus  $P_i$.  We used the approximate functional relations revealed by these data to estimate the transient reactance and  the inertia constant for each generator in the Guatemala, Northern Italy, and Poland systems. 
}
\end{center}
\end{figure*}

\clearpage

\newpage
\baselineskip12pt
\setlength{\oddsidemargin}{-0.35in}
\setlength{\evensidemargin}{\oddsidemargin}
\setlength{\hsize}{7.2in}
{\sffamily
\subsubsection*{\label{table1}
\sffamily\small 
Table 1: Structural properties and synchronization stability of the systems considered.}
\vspace{-5mm}
{\scriptsize
\begin{tabular*}{\hsize}{@{\extracolsep{\fill}}lccccccccccc}
\hline\hline \rule[4mm]{0mm}{0mm} 
& & & & \multicolumn{3}{c}{System heterogeneity$^\dag$} 
& \multicolumn{4}{c}{Synchronization stability$^\ddag$}\\
\cline{5-7}\cline{8-11}
\rule[4mm]{0mm}{0mm}System$^\S$
& Nodes & Links & Generators 
& {\rmfamily\textbf{Y}$_0$} & {\rmfamily\textbf{Y}} & $\beta_i$
& \multicolumn{1}{c}{Original} 
& \multicolumn{1}{c}{$x'_{d,i}$ adjusted}
& \multicolumn{1}{c}{$\beta_i = \bar\beta$}
& \multicolumn{1}{c}{$\beta_i=\beta_\text{opt}\!\!\!\!\!$}\\
\hline
\rule{0pt}{8pt}3-generator test system & 9 & 9 & 3 &   0.39 &   0.09 &   0.83 &  -1.71 &---&  -2.21 &  -8.69\\
10-generator test system & 39 & 46 & 10 &   0.81 &   0.38 &   0.37 &  -0.24 &---&  -0.37 &  -3.65\\
50-generator test system & 145 & 422 & 50 &   1.91 &   1.21 &   2.07 &  -0.02 &---&  -1.53 &  -1.75\\
Guatemala power grid & 370 & 392 & 94 &   6.86 &   1.18 &   1.22 &  -0.33 &---&  -0.02 &  -4.19\\
Northern Italy power grid  & 678 & 822 & 170 &   3.42 &   0.84 &   2.02 &   \rule{1mm}{0mm}7.20 &  -0.48 &  -1.59&  -3.70\\
Poland power grid  & 2383 & 2886 & 327 &   2.26 &   3.00 &   0.92 & 140.07\rule{2mm}{0mm} &  -0.03 &  -0.02&  -1.53\\
\hline\hline
\end{tabular*}

\vspace{-2mm}\noindent
$^\dag$As a measure of heterogeneity,  we 
show the standard deviation normalized by the average. 
The quantities considered are the weighted degrees computed for {\rmfamily\textbf{Y}$_0$} and {\rmfamily\textbf{Y}}, which represent the structure of the physical
and effective network, respectively, and  the parameter $\beta_i$, which represents properties of the generators.
$^\ddag${As a measure of synchronization stability, we show the Lyapunov exponent $\lambda^{\max}$.
We consider this exponent for the original parameter values, for the transient reactance $x'_{d,i}$ adjusted, for  
all $\beta_i$ set equal to their average $\bar\beta$, 
and for the generator parameter $R_i$ and/or $D_i$ adjusted to ensure $\beta_i = \beta_\text{opt}$.}
$^\S${See Methods for a description of data sources.}
}
}

\newcommand{\vs}{-1mm}
\newcommand{\vstwo}{-2mm}

\newpage
\setlength{\oddsidemargin}{0in}
\setlength{\evensidemargin}{\oddsidemargin}
\setlength{\hsize}{15cm}
\baselineskip12pt

{\sffamily
\subsubsection*{\noindent
\sffamily\small \vspace{-4mm}
\!\!\!Table 2: Phase difference in synchronous states and the real versus imaginary parts of the admittances in the physical and effective networks.}
\vspace{-5mm}
\sffamily\scriptsize\mdseries
\begin{tabular*}{\hsize}{@{\extracolsep{\fill}}lrrrcc}
\\[-5mm]
\hline
\hline
& & \multicolumn{2}{c}{\rule[-5pt]{0pt}{15pt}Physical network$^\dag$}
& \multicolumn{2}{c}{Effective network} \\ 
\cline{3-4}\cline{5-6}
System
& Mean $|\delta_{ij}^*|/\pi$
& \multicolumn{1}{c}{\rule[-5pt]{0pt}{15pt} Mean $|G_{0ij}|$} & \multicolumn{1}{c}{Mean $|B_{0ij}|$}
& \multicolumn{1}{c}{\rule[-5pt]{0pt}{15pt} Mean $|G_{ij}|$} & \multicolumn{1}{c}{Mean $|B_{ij}|$} \\
\hline
\rule{0pt}{10pt}3-generator system & 0.06\rule{5mm}{0mm} & 0.95 \rule{3mm}{0mm} &  12.0 \rule{3mm}{0mm} & 0.2367 &  1.275\\[\vs]
10-generator system & 0.07\rule{5mm}{0mm} & 6.81 \rule{3mm}{0mm} &  80.9 \rule{3mm}{0mm} & 0.3784 &  1.036\\[\vs]
50-generator system & 0.12\rule{5mm}{0mm} & 2.57 \rule{3mm}{0mm} &  46.3 \rule{3mm}{0mm} & 0.1978 &  1.242\\[\vs]
Guatemala & 0.08\rule{5mm}{0mm} & 7.74 \rule{3mm}{0mm} &  62.1 \rule{3mm}{0mm} & 0.0015 &  0.004\\[\vs]
Northern Italy: \hspace{0mm}original & 0.07\rule{5mm}{0mm} & 105.88 \rule{3mm}{0mm} & 758.3 \rule{3mm}{0mm} & 0.0106  &  0.039\\[\vstwo]
\phantom{Northern Italy: \hspace{0mm}}$x'_{d,i}$ adjusted & 0.05\rule{5mm}{0mm} &  &  & 0.0104 &  0.042\\[\vs]
Poland: \hspace{8.5mm}original & 0.25\rule{5mm}{0mm} & 24.06 \rule{3mm}{0mm} & 597.3 \rule{3mm}{0mm} & 0.0012 &  0.018\\[\vstwo]
\rule[-5pt]{0pt}{5pt}\phantom{Poland: \hspace{8.5mm}}$x'_{d,i}$ adjusted & 0.13\rule{5mm}{0mm} & &  & 0.0013 &  0.019\\
\hline\hline
\end{tabular*}
\\[1mm]
\scriptsize
\baselineskip12pt
$^\dag$The real and imaginary components of the admittance $-Y_{0ij}$ 
are denoted $-G_{0ij}$ and $-B_{0ij}$, respectively.
}

\vspace{-4mm}

\setlength{\hsize}{9cm}
{\sffamily
\subsubsection*{\noindent
\sffamily\small \vspace{-4mm}
\!\!Table 3: 2-norm of the symmetric  and antisymmetric parts of the matrix $\textrm{\textbf{P}}''$.}
\sffamily\scriptsize\mdseries
\begin{tabular*}{\hsize}{@{\extracolsep{\fill}}lrc}
\\[-10mm]
\hline\hline
\rule[-4pt]{0pt}{14pt}System & Symmetric & Antisymmetric \\
\hline
\rule{0pt}{10pt}3-generator system &   0.68\rule{3mm}{0mm} &  0.011\\[\vs]
10-generator system &   0.48\rule{3mm}{0mm} &  0.018\\[\vs]
50-generator system &   1.89\rule{3mm}{0mm} &  0.016\\[\vs]
Guatemala &   3.04\rule{3mm}{0mm} &  0.085\\[\vs]
Northern Italy: \hspace{0mm}original &   2.45\rule{3mm}{0mm} &  0.030\\[\vstwo]
\phantom{Northern Italy: \hspace{0mm}}$x'_{d,i}$ adjusted &  34.27\rule{3mm}{0mm} &  0.068\\[\vs]
Poland: \hspace{8.5mm}original &  14.45\rule{3mm}{0mm} &  0.261\\[\vstwo]
\rule[-5pt]{0pt}{5pt}\phantom{Poland: \hspace{8.5mm}}$x'_{d,i}$ adjusted & 234.24\rule{3mm}{0mm} &  0.982\\\hline\hline
\end{tabular*}
}

\clearpage

\setlength{\hsize}{10cm}
{\sffamily
\subsubsection*{\noindent
\sffamily\small \vspace{-4mm}
\!\!\!Table 4: Fraction of positive off-diagonal elements in the matrices $\boldsymbol{(B_{ij})}$, $\boldsymbol{(\cos\delta^*_{ij})}$, and  $\boldsymbol{(B_{ij}\cos\delta^*_{ij})}$.}
\sffamily\scriptsize\mdseries
\begin{tabular*}{\hsize}{@{\extracolsep{\fill}}lccc}
\\[-10mm]
\hline\hline
\rule[-4pt]{0pt}{14pt}System & $B_{ij}$ & $\cos\delta^*_{ij}$ & $B_{ij}\cos\delta^*_{ij}$\\
\hline
\rule{0pt}{10pt}3-generator system &  1.000 &  1.000 &  1.000\\[\vs]
10-generator system &  1.000 &  1.000 &  1.000\\[\vs]
50-generator system &  0.689 &  0.995 &  0.687\\[\vs]
Guatemala &  0.999 &  1.000 &  0.999\\[\vs]
Northern Italy: \hspace{0mm}original &  0.999 &  0.973 &  0.973\\[\vstwo]
\phantom{Northern Italy: \hspace{0mm}}$x'_{d,i}$ adjusted &  0.997 &  1.000 &  0.997\\[\vs]
Poland: \hspace{8.5mm}original &  0.978 &  0.829 &  0.810\\[\vstwo]
\rule[-5pt]{0pt}{5pt}\phantom{Poland: \hspace{8.5mm}}$x'_{d,i}$ adjusted &  0.957 &  0.992 &  0.949\\
\hline\hline
\end{tabular*}
}

\end{document}